\documentclass[dvips,12pt,a4paper]{article}
\usepackage{a4wide}
\usepackage{epsfig}
\usepackage{amsmath}
\usepackage{latexsym}

  \newlength{\absize}
\newcommand{\dd}{\mbox{{\rm d}}}

\newcommand{\third}{{\textstyle\frac{1}{3}}}

\renewcommand{\Re}{\mbox{Re}}

\newcommand{\Lumint}{{\cal L}_{\rm int}}
\def \sup{^{\vphantom{2}}}
\begin{document}
  \thispagestyle{empty}
  \pagestyle{empty}
  \renewcommand{\thefootnote}{\fnsymbol{footnote}}
\newpage\normalsize
    \pagestyle{plain}
    \setlength{\baselineskip}{4ex}\par
    \setcounter{footnote}{0}
    \renewcommand{\thefootnote}{\arabic{footnote}}
\newcommand{\preprint}[1]{%
  \begin{flushright}
    \setlength{\baselineskip}{3ex} #1
  \end{flushright}}
\renewcommand{\title}[1]{%
  \begin{center}
    \LARGE #1
  \end{center}\par}
\renewcommand{\author}[1]{%
  \vspace{2ex}
  {\Large
   \begin{center}
     \setlength{\baselineskip}{3ex} #1 \par
   \end{center}}}
\renewcommand{\thanks}[1]{\footnote{#1}}
\renewcommand{\abstract}[1]{%
  \vspace{2ex}
  \normalsize
  \begin{center}
    \centerline{\bf Abstract}\par
    \vspace{2ex}
    \parbox{\absize}{#1\setlength{\baselineskip}{2.5ex}\par}
  \end{center}}

\title{Optimal observables for new-physics search at LEP2}
\vfill
\author{P. Osland and A.A. Pankov\footnote{Permanent address: Gomel
Polytechnical Institute, Gomel, 246746 Belarus.} \\\hfil\\
        Department of Physics\thanks{Electronic mail addresses:
                {\tt Per.Osland@fi.uib.no; pankov@gpi.gomel.by}}\\
        University of Bergen \\ All\'egt.~55, N-5007 Bergen, Norway}
                                                                       
\vfill
\abstract
{New observables $\sigma_\pm$ for the process
$e^+e^-\to\mu^+\mu^- $ allow one to get more direct information
on additional $Z'$ boson effects than what is
obtained from the canonical ones, $\sigma$ and $A_{\rm FB}$.
Their deviations from the Standard Model predictions 
have very specific energy dependences, 
which are precisely determined by SM parameters.
At energies varying from TRISTAN to LEP2, 
one can uniquely predict the signs of $\Delta\sigma_{\pm}$ 
induced by a $Z'$ 
as well as the locations of their extrema and zeros.
This unambiguous energy correlation
could be quite useful in distinguishing effects due to $Z'$ exchange
from those caused by other new physics sources.
Furthermore, there are two energy points, $\sqrt{s_{+}}\simeq$ 78 GeV
and $\sqrt{s_{-}}\simeq$ 113 GeV, where
the SM quantities $\sigma_\pm^{\rm SM}$ as well as the deviations
$\Delta\sigma_{\pm}$ attain their minimum values or vanish.
These points could be very favourable for a search for new
physics beyond the SM and beyond $Z'$ effects.

}
\newpage
    \setcounter{footnote}{0}
    \renewcommand{\thefootnote}{\arabic{footnote}}
    \setcounter{page}{1}
\section{Introduction}
After several years of successful LEP1 operation with high statistics, 
there is excellent agreement between the data and the Standard Model (SM) 
predictions at energies around the $Z$ resonance \cite{Warsaw, altarelli}.
In addition, these experiments at LEP1 have tested the SM expectations
away from the $Z$ pole. In particular, experimental 
results from studies of events collected
in the channel $e^+e^-\to\mu^+\mu^-\gamma_{\rm isr}$ \cite{lep1-rad},
with $\gamma_{\rm isr}$ being an initial-state radiation
photon, are at LEP1 used to probe the cross section and 
forward-backward asymmetry in the energy region between LEP1 and TRISTAN 
and down to PETRA energies.

The investigation of the process $e^+e^-\to\mu^+\mu^-$ 
at energies below the $Z$ peak is attractive
because preliminary results from TRISTAN \cite{kek} could
indicate a downward deviation of the cross section 
(by two standard deviations) from the prediction
of the SM at 58~GeV.
Measurable deviations in the $e^+e^-\to\mu^+\mu^-$ cross section
are in this energy range predicted by several models beyond the SM,
for instance those which introduce an additional $Z'$ boson
\cite{p1}.  Alternatively, it could be induced by 
anomalous triple gauge boson couplings \cite{lep2}. 
Thus, $Z'$ effects at $\sqrt{s}<M_Z$, as well as
at LEP2 would be of a similar type as those arising from 
anomalous triple gauge couplings, although the responsible mechanism 
would be of a totally different origin.

It is very important to optimize the strategy when searching for new
physics beyond the SM, since any signal would
most likely be very small.
It is also important to exploit the available data at ``low''-energy
machines,  namely TRISTAN \cite{kek} and LEP1 \cite{lep1-rad},
as well as those at LEP1.5 \cite{lep1.5}.
We concentrate here on the strategy for new-physics search
at these machines, in particular at LEP2, where data will be obtained
in the next couple of years.

In general, one is not able to predict the magnitude of the $Z'$
effects because they depend on {\it a priori} unknown parameters:
the couplings to the $Z'$ and its mass, $M_{Z'}$.
Therefore, definite predictions of $Z'$ effects would be quite 
desirable and important in such searches, also in order to 
discriminate them from other possible new physics effects.

In a previous paper \cite{paper1} we studied the interference effects
induced by an extra neutral gauge boson $Z'$ in the production of
lepton pairs 
\begin{equation}
\label{Eq:leptons}
e^+e^-\to l^+l^-, \qquad \mbox{($l=\mu$, $\tau$)}.
\end{equation}
We have shown that assuming lepton universality, the lepton channel has 
the advantage over the $q\bar q$ channel
that the signs of the interference terms are given very simply by the
propagators of $Z$ and $Z'$. This is caused by the fact that
the observables $\sigma$ and $A_{\rm FB}$
depend only on {\it squares} of coupling constants.
Due to this dependence, the canonical observables have for the
process (\ref{Eq:leptons}) certain properties which are useful for
the identification of effects of $Z'$ origin.
Namely, at LEP2 energies, the effect of a $Z'$ (with arbitrary
vector and axial vector couplings) is to {\it reduce both the
cross section and the forward-backward cross-section difference}
\cite{paper1}, as compared with the SM expectation.
These unique properties of $\sigma$ and $\sigma_{\rm FB}$ are due to the
fact that the $\gamma-Z'$, as well as the $Z-Z'$ interference terms are
both negative. However, predictions for another energy region, 
$\sqrt{s}<M_Z$, are less definite.
For example, at energies below the $M_Z$ the modifications 
of the cross section and the forward-backward asymmetry depend crucially
on whether the coupling is dominantly vector or dominantly axial vector.
In such a situation, it may be quite difficult to uniquely identify and
extract effects due to extra gauge bosons from those caused by other
new physics effects.

In this paper we extend the analysis started in \cite{paper1}. 
As we shall show below it is possible to provide more definite information on
$Z'$ effects. We shall here consider certain new observables, for which
the deviations from the SM predictions
have very specific energy dependences.
These energy dependences are precisely determined because 
they involve only the SM parameters such as the lepton 
couplings of the standard $Z$ boson and the mass $M_Z$. 
In particular, one can uniquely predict the sign of any deviations of the 
observables due to a $Z'$ at energies from TRISTAN to LEP2, as well as  
the locations of their extrema and zeros. In such a case, one can easily 
distinguish the effects induced by $Z'$ from those caused by other new 
physics effects.

\section{New observables}
A new neutral gauge boson would induce additional neutral current 
interactions, the corresponding Lagrangian can be written as
\begin{equation}
-{\cal L}_{\rm NC}=eJ^{\mu}_{\gamma}A_{\mu}+
g_ZJ^{\mu}_ZZ_{\mu}+g_{Z'}J^{\mu}_{Z'}Z'_{\mu},
\end{equation}
where $e=\sqrt{4\pi\alpha}$, $g_Z=e/s_Wc_W$ 
($s^2_W=1-c^2_W\equiv\sin^2\theta_W$) and $g_{Z'}$ are the gauge coupling
constants. The neutral currents are
\begin{equation}
J^\mu_i=\sum_{f}\bar\psi_f\gamma^\mu\left(L^f_iP_L+R^f_iP_R\right)\psi_f=
\sum_{f}\bar\psi_f\gamma^\mu\left(V^f_i-A^f_i\gamma_5\right)\psi_f,
\label{current}
\end{equation}
where $i\equiv\gamma$, $Z$, $Z'$, and $P_{L,R}=(1\mp\gamma_5)/2$ are the 
left- and right-handed chirality projection operators. 
The SM vector and axial-vector couplings of the vector boson $i$ to the
fermions are
\begin{equation}
V^f_\gamma=Q_f, \qquad A^f_\gamma=0, \qquad
V^f_Z=\frac{I^f_{3L}}{2}-Q_fs_W^2, \qquad A^f_Z=\frac{I^f_{3L}}{2}.
\end{equation}
Here, $Q_f$ is the electric charge of $f$ ($Q_e=-1$),
and $I^f_{3L}$ denotes the third component of the weak isospin.

The lowest-order unpolarized differential cross section for the process
(\ref{Eq:leptons}), assuming $e$-$l$ universality, mediated by
$\gamma$, $Z$, and the extra $Z'$ boson exchanges, is given by
\begin{eqnarray}
\label{Eq:dsigma}
\frac{\dd\sigma}{\dd\cos\theta}
&=&\frac{\pi\alpha^2}{2s}
\left[(1+\cos^2\theta)\, F_1 +2\cos\theta\, F_2\right], \nonumber \\[4mm]
F_1&=&F_1^{\rm SM}+\Delta F_1, \qquad F_2=F_2^{\rm SM}+\Delta F_2,
\end{eqnarray}
with ($v\equiv v_l$, $a\equiv a_l$ and similarly for the primed
quantities):
\begin{eqnarray}
F_1^{\rm SM}&=&1+2\,v^2\Re\chi\sup_Z
+(v^2+a^2)^2|\chi\sup_Z|^2,\nonumber \\[2mm]
F_2^{\rm SM}&=&2\,a^2\Re\chi\sup_Z
+4\,(va)^2|\chi\sup_Z|^2, \nonumber \\[2mm]
\Delta F_1&=&2\,v'{}^2\Re\chi\sup_{Z'}
+2\,(vv'+aa')^2\Re(\chi\sup_{Z}\chi^*_{Z'})
+(v'{}^2+a'{}^2)^2|\chi\sup_{Z'}|^2, \nonumber \\[2mm]
\Delta F_2&=&2\,a'{}^2\Re\chi\sup_{Z'}
+2(va'+v'a)^2\Re(\chi\sup_{Z}\chi^*_{Z'})
+4\,(v'a')^2|\chi\sup_{Z'}|^2.
\end{eqnarray}
The coupling constants are 
normalized to the unit of charge $e$, and are expressed in terms of the 
couplings in the current basis (\ref{current}) as
\begin{eqnarray}
\label{Eq:couplings}
v&=&\frac{g_Z}{e}V^f_Z=\frac{1}{4 s_W c_W}(-1 +4 s^2_W),
\qquad
a=\frac{g_Z}{e}A^f_Z=\frac{-1}{4 s_W c_W}, \nonumber \\
v'&=&\frac{g_{Z'}}{e}V^f_{Z'},
\qquad
a'=\frac{g_{Z'}}{e}A^f_{Z'},
\end{eqnarray}
and the gauge boson propagators are
$\chi\sup_V=s/(s-M^2_V+iM_V\Gamma_V)$, $V=Z$, $Z'$.
The total cross section and the forward-backward asymmetry 
can be written as 
\begin{equation}
\label{Eq:AFB}
\sigma=\sigma_{\rm pt}\, F_1, \qquad
A_{\rm FB}=\frac{\sigma_{\rm F}-\sigma_{\rm B}}
{\sigma_{\rm F}+\sigma_{\rm B}}
=\frac{3F_2}{4F_1}\ ,
\end{equation}
with $\sigma_{\rm pt}=(4\pi\alpha^2)/(3s)$.

In the search for effects induced by the exchange of a $Z'$,
it will be advantageous to consider new observables, free of certain
shortcomings of the canonical ones, $\sigma$ and $A_{\rm FB}$.
The observables we want to propose, are differences of cross sections
obtained by integrating over suitable ranges of polar angle,
\begin{eqnarray}
\label{sigma+}
\sigma_+&\equiv&\left(\int_{-z^*}^1-\int_{-1}^{-z^*}\right)
\frac{\dd\sigma}{\dd\cos\theta}\, \dd\cos\theta, \\
\label{sigma-}
\sigma_-&\equiv&\left(\int_{-1}^{z^*}-\int_{z^*}^1\right)
\frac{\dd\sigma}{\dd\cos\theta}\, \dd\cos\theta, 
\end{eqnarray}
where $z^*>0$ is determined from the condition that the coefficients
multiplying $F_1$ and $F_2$ be the same
(cf. Eq.~(\ref{Eq:dsigma})),
\begin{equation}
\int_{-z^*}^{z^*}(1+\cos^2\theta)\,\dd\cos\theta
=\left(\int_{z^*}^1-\int_{-1}^{-z^*}\right)\, 2\cos\theta\, \dd\cos\theta,
\end{equation}
or 
\begin{equation}
\third z^*{}^3+z^*{}^2+z^*-1=0,
\end{equation}
whose solution is $z^*=2^{2/3}-1=0.5874$, corresponding to
$\theta^*=54^\circ$.
Thus,
\begin{equation}
\label{Eq:sigmapm}
\sigma_{\pm}=\sigma^*_{\rm pt}\,(F_1\pm F_2),
\end{equation}
where
\begin{equation}
\sigma^*_{\rm pt}
=\frac{3}{4}\left(1-z^*{}^2\right)\sigma_{\rm pt}
=\frac{\pi\alpha^2}{s}\left(1-z^*{}^2\right).
\end{equation}

It should be noted, that the introduction and exploitation of the new 
independent observables $\sigma_{\pm}$ is quite analogous to dealing with
the canonical ones, $\sigma$ and $A_{\rm FB}$. 
In fact, from Eqs.~(\ref{Eq:AFB}) and (\ref{Eq:sigmapm}) 
one can simply express $\sigma_{\pm}$ in terms of 
$\sigma$ and $A_{\rm FB}$\footnote{The proposed observables are
related to helicity amplitudes as follows:
$\sigma_+\propto|A_{\rm RR}|^2+|A_{\rm LL}|^2$;
$\sigma_-\propto|A_{\rm RL}|^2+|A_{\rm LR}|^2$.}:
\begin{equation}
\label{relation}
\sigma_{\pm}=\frac{3}{4}\left(1-z^*{}^2\right)\sigma
\left(1\pm\frac{4}{3}A_{\rm FB}\right)=
0.49\,\sigma\left(1\pm\frac{4}{3}A_{\rm FB}\right).
\end{equation}
Thus, they can be measured either directly according to Eqs.
(\ref{sigma+}) and (\ref{sigma-}), or indirectly by means of $\sigma$ and 
$A_{\rm FB}$\footnote{It means that the available experimental data
for $\sigma$ and $A_{\rm FB}$ of the process (\ref{Eq:leptons})
at TRISTAN to LEP1.5 energies \cite{lep1-rad, kek, lep1.5}
can be directly converted to $\sigma_\pm$.}.

For the sake of a simplified presentation,
the discussion presented in the next section is 
based on several assumptions, whereas our numerical results are 
based on the full formulas. 
In particular, we assume: \\ \noindent
({\it i}) since the typical upper bound for the $Z^\prime$ boson mass,
$M_{Z^\prime}>600$ GeV \cite{Tevatron}, 
lies quite a bit higher than the energy available at LEP2, 
it suffices to take into account $Z^\prime$ interference effects
only, the pure $Z^\prime$ exchange contributions being negligible;
\\ \noindent
({\it ii}) since in the SM 
$\vert v\vert\ll\vert a\vert<1$, in the following we shall ignore
$v$ against $a$. 
In addition, one can neglect the imaginary part of 
the $Z^\prime$ boson propagator.

According to these simplifying assumptions,
we can write
\begin{eqnarray}
\label{Eq:F1F2}
\sigma_{\pm}/\sigma^*_{\rm pt}=
F_1\pm F_2
&=&\left(F_1^{\rm SM}\pm F_2^{\rm SM}\right)
+\left(\Delta F_1\pm \Delta F_2\right) \nonumber \\
&\underset{\sqrt{s}\ll M_{Z'}}{\approx}&|1\pm a^2\chi\sup_{Z}|^2
+2(v'{}^2\pm a'{}^2)\chi\sup_{Z'}(1\pm a^2\, \Re\,\chi\sup_{Z}),
\end{eqnarray}
where the first term represents the SM contribution, and the
second one the $Z'$ effects.

\section{Improved Born results}

The previous formula for the differential cross section (\ref{Eq:dsigma})
as well as those for all other observables
are still valid to a very good (improved Born) approximation after 
one-loop electro-weak radiative corrections, with the following
replacements \cite{Altarelli1}:
\begin{eqnarray}
\label{impborn}
\alpha&\Rightarrow&\alpha(M_Z^2) \nonumber
\\
v&\Rightarrow&
\frac{1}{\sqrt{\kappa}}
\left(I^e_{3L}-2Q_e\sin^2\theta^{\rm eff}_W\right), \qquad
a\Rightarrow
\frac{I^e_{3L}}{\sqrt{\kappa}}
\nonumber \\
\sin^2\theta_W
&\Rightarrow&
\sin^2\theta^{\rm eff}_W, \qquad
\sin^2(2\theta^{\rm eff}_W)
\equiv\kappa=\frac{4\pi\alpha(M_Z^2)}{\sqrt{2}G_{\rm F}\, M_Z^2\rho},
\end{eqnarray}
with
\begin{equation}
\label{rho}
\rho\approx 1+\frac{3 G_{\rm F} m^2_t}{8\pi^2\sqrt{2}},
\end{equation}
where only the main contribution to $\rho$, coming from the top mass, 
has been given. This parameterization uses the best known quantities
$G_{\rm F}$, $M_Z$, and $\alpha(M^2_Z)$.
A final step consists in introducing the energy dependence in
the width term of the $Z$ propagator,
\begin{equation}
\label{prop}
\chi\sup_Z(s)\Rightarrow\frac{s}{s-M_Z^2+i(s/M_Z^2)M_Z\Gamma_Z}.
\end{equation}
All numerical results presented in this section are based on
the improved Born approximation with $m_t=170$~GeV and
$m_H=300$~GeV.

Let us start our discussion with the observable $\sigma_+$, defined 
by Eqs.~(\ref{Eq:sigmapm}) and (\ref{Eq:F1F2}). It has an SM part,
$|1+a^2\chi\sup_{Z}|^2$ which tends quadratically to its minimum
value (given by the $Z$ width) at\footnote{This special energy,
as well as the one given by eq.~(\ref{Eq:smin}),
have also been noted by Fr\`ere et al.\ \cite{frere}
from the study of helicity cross sections.}
\begin{equation}
\label{Eq:splus}
\sqrt{s_+}=\frac{M_Z}{\sqrt{1+a^2}}\simeq78~\mbox{GeV},
\end{equation}
as is displayed in Fig.~1a.
While the $Z'$ interference term also vanishes at the same  point 
(see Fig.~1b), it does so only linearly. Therefore, one may expect 
an enhanced sensitivity of $\sigma_+$ to $Z'$ effects around this point, 
$\sqrt{s_+}$. (In Figs.~1 and 2 we consider, as illustrative cases, 
the effects of a $Z'$ with $M_{Z'}=600$~GeV and various
leptonic couplings.)

From Eqs.~(\ref{Eq:sigmapm}) and (\ref{Eq:F1F2}) one can directly read off
the deviations of $\sigma_+$ from the SM prediction at ``low'' energies 
\begin{eqnarray}
\label{Eq:Deltasig}
\Delta\sigma_{+}&\equiv&\sigma_{+}-\sigma_{+}^{\rm SM} \nonumber \\
&\approx&
2\,\sigma^*_{\rm pt}\,
(v'{}^2+ a'{}^2)\chi\sup_{Z'}(1+ a^2\, \Re\,\chi\sup_{Z}).
\end{eqnarray}
Before embarking on a more detailed analysis, two important remarks 
are in order. First, as one can see from Eq.~(\ref{Eq:Deltasig}),
the dependence of $\Delta\sigma_{+}$ on the $Z'$ parameters 
is characterized by the expression $(v'{}^2+ a'{}^2)\chi\sup_{Z'}$, 
where the $Z'$ couplings appear only as a sum of their squares,
i.e., as a positive definite quantity.
This means that in contrast to the canonical observables, 
$\sigma$ and $A_{\rm FB}$ \cite{paper1}, in $\sigma_{+}$ 
the $\gamma-Z'$ and $Z-Z'$ interferences contribute coherently.
Thus, {\it in $\Delta\sigma_{+}$, there is no cancellation between the
$\gamma-Z'$ and $Z-Z'$ interference effects}, instead they enhance 
each other.

Secondly, the energy dependence of $\Delta\sigma_{+}$ is given
by the factor $(1+ a^2\, \Re\,\chi\sup_{Z})$, which is
completely determined by Standard-Model parameters. 
Hence, one can precisely predict the {\it energy dependence} of 
the deviation $\Delta\sigma_{+}$. Its important feature is that it is
independent of the $Z'$ lepton couplings and the mass $M_{Z'}$. 
In particular, $\Delta\sigma_{+}$ vanishes 
at $\sqrt{s}=\sqrt{s_+}$ 
and $\sqrt{s}\approx M_Z$, and it achieves extrema at 
\begin{equation}
\label{Eq:s+}
\vert\sqrt{s_+}-\sqrt{s}\vert\approx\frac{\Gamma_Z}{2}.
\end{equation}

It should be stressed once again that in the approximation
$v=0$ the locations of these particular points (zeros and extrema) 
are not affected by a variation of the $Z'$ 
parameters. The finite value of $v$ leads to a small
shift, by a factor $1+\delta$, where $\delta=va\cos\gamma/(1+a^2)$,
with $\cos\gamma=2(v'/a')/[1+(v'/a')^2]$. This amounts to a shift
of at most (for $|\cos\gamma|=1$) 1.6~GeV. 
Also, the non-zero value of $v$ is responsible for the splitting
of the three curves in Fig.~1a, for different values of $v'a'$.

Fig.~1b shows the energy dependence of the relative deviation
\begin{equation}
\frac{\Delta\sigma_+}{\sigma_+^{\rm SM}}\equiv
\frac{\sigma_+-\sigma_+^{\rm SM}}{\sigma_+^{\rm SM}},
\label{deltasig}
\end{equation}
for different couplings: $v'{}^2+a'{}^2=0.25$, 0.5, and 1. 
One can see from Fig.~1b that the $Z'$ interference pattern
is quite stable under variations of the couplings, only its scale is
changed.

Another very important property of the observable $\sigma_+$
is that the sign of the deviation $\Delta\sigma_+$ is uniquely determined,
cf.\ Eq.~(\ref{Eq:Deltasig}). In fact, at $\sqrt{s}<\sqrt{s_+}$ and
$\sqrt{s}> M_Z$ it is negative ($\Delta\sigma_+< 0$), while for
$\sqrt{s_+}<\sqrt{s}< M_Z$ the quantity $\Delta\sigma_+> 0$ (see Fig.~1b).
In other words, there is a correlation of the signs of $\Delta\sigma_+$ 
at different energies. This means that the determination of the signs 
of the deviation $\Delta\sigma_+$ at different energies should help
in distinguishing $Z'$ effects from those induced by other possible 
new physics origins. 
For example, if at $\sqrt{s}<\sqrt{s_+}$ or at $\sqrt{s}>M_Z$ one observes 
$\Delta\sigma_+>0$, then one can definitely conclude that it is not 
induced by a $Z'$. However, if $\Delta\sigma_+$ is negative at 
$\sqrt{s}<\sqrt{s_+}$ or/and at $\sqrt{s}>M_Z$,
and positive for $\sqrt{s_+}<\sqrt{s}< M_Z$,
then one has a stronger case that it could be due to a $Z'$ (see Fig.~1b).

Finally, another interesting feature of the energy dependence of 
$\Delta\sigma_+$ is associated with the energy point $\sqrt{s_+}$. As 
mentioned above, at this energy the SM background ($\sigma_+^{\rm SM}$) 
tends to its minimum value, and also $\Delta\sigma_+$ vanishes. 
Thus, this energy is very favourable for a search for new physics 
{\it beyond the SM and beyond $Z'$ effects.}

In Fig.~1c we show the statistical significance 
\begin{equation}
S_+\equiv\frac{|\sigma_{+}-\sigma_{+}^{\rm SM}|}
{\delta\sigma_{+}}
=\frac{|\Delta\sigma_{+}|}
{\sqrt{\sigma^{\rm SM}}}\,
\sqrt{\Lumint}\, ,
\label{SS+}
\end{equation}
defined as deviation from the SM prediction in units of the standard 
deviation, where $\delta\sigma_{+}$ is the statistical uncertainty,
and $\Lumint$ the integrated luminosity, $\Lumint=\int\dd t{\cal L}$.
We here consider 
$\Lumint=300~\mbox{pb}^{-1}$ (as a typical value at TRISTAN), and  
$M_{Z'}=600$~GeV.
As can be seen from the figure, 
with the exception of the two minima, this function
increases with $\sqrt{s}$.
These features are quite independent of the $Z'$ mass.
We note that $S_+$ has a minimum around $\sqrt{s_+}$.
Hence, this energy is very favourable for a search for new physics 
other than $Z'$ effects.

Let us now turn to the observable $\sigma_-$ defined by 
Eqs.~(\ref{Eq:sigmapm}) and (\ref{Eq:F1F2}). 
It has several properties in common with $\sigma_+$.
In particular, the deviation from the SM prediction,
\begin{eqnarray}
\label{Eq:Dsigm}
\Delta\sigma_{-}&\equiv&\sigma_{-}-\sigma_{-}^{\rm SM} \nonumber \\
&=&
2\,\sigma^*_{\rm pt}\,
(v'{}^2- a'{}^2)\chi\sup_{Z'}(1- a^2\, \Re\,\chi\sup_{Z}).
\end{eqnarray}
has an energy dependence given by SM parameters.
However, the sign and magnitude of such a deviation is determined by 
$v'{}^2-a'{}^2$, or whether the $Z'$ couplings are predominantly
vector or axial vector\footnote{One should note that the $\sigma_{-}$
dependence on $(v'{}^2-a'{}^2)$ is exact, 
there is no dependence on $v'a'$.}. 
Thus, this deviation, at a given energy,
can be either positive or negative. 

Furthermore, the energy where 
$\sigma_-^{\rm SM}$ has its minimum and $\Delta\sigma_-=0$ 
is located above $M_Z$, rather than below, namely at
\begin{equation}
\label{Eq:smin}
\sqrt{s_-}=\frac{M_Z}{\sqrt{1-a^2}}\simeq 113~\mbox{GeV}.
\end{equation}
The corresponding set of energy correlations of a $Z'$ signal in 
$\sigma_-$ is given in Table 1.
\begin{table}
\begin{center}
\begin{tabular}{||l|c|c|c||}                     \hline
$\sqrt{s}$ &$<M_Z$ & $M_Z$--113 GeV & $>113$ GeV \\ \hline
& $-$ & + & $-$ \\ \cline{2-4} 
\raisebox{1.5ex}[0cm][0cm]{sign of $\Delta\sigma_-$} 
& + & $-$ & + \\ \hline
\end{tabular}
\end{center}
\caption{Energy correlation of a $Z'$ signal in $\sigma_-$.}
\end{table}
Finally, we note that due to the different dependences on 
the $Z'$ couplings ($v'{}^2+a'{}^2$ vs. $v'{}^2-a'{}^2$),
the observables $\sigma_+$ and $\sigma_-$ are quite complementary 
in the $Z'$ search.

The energy dependences of the observable $\sigma_-$, 
its relative deviation from the SM prediction, 
\begin{equation}
\frac{\Delta\sigma_-}{\sigma_-^{\rm SM}}\equiv
\frac{\sigma_{-}-\sigma_-^{\rm SM}}{\sigma_-^{\rm SM}},
\label{deltasigmin}
\end{equation}
and the statistical significance, defined as 
\begin{equation}
S_-\equiv\frac{|\sigma_{-}-\sigma_{-}^{\rm SM}|}
{\delta\sigma_{-}}\, ,
\label{SS-}
\end{equation}
are shown in Figs.~2b and 2c, respectively.
In particular, in Fig.~2b  we show the
relative deviation $\Delta\sigma_-/\sigma_-^{\rm SM}$ for
two different possibilities of the parameters,
$({v'}^2-{a'}^2)>0$ and $({v'}^2-{a'}^2)<0$,
henceforth denoted as the $V$ and $A$ case, respectively. As can be seen 
from Fig.~2c the $A$ case provides somewhat higher sensitivity to a $Z'$
than the $V$ case at $\sqrt{s}>M_Z$. 
The reason is that in the $V$ case there is at high energies 
some cancellation between the interference terms and the pure $Z'$ 
contribution.

Finally, the energy region around $\sqrt{s_-}\simeq113$~GeV 
where $\sigma_{-}^{\rm SM}$ has its minimum value, 
and where $\Delta\sigma_{-}$ also vanishes, 
would presumably be a convenient place
to probe for new physics effects beyond those due to a $Z'$ (see Fig.~2c).

\section{Model-independent bounds on $Z'$ couplings}
In this section we assess the sensitivities of the observables
$\sigma_\pm$ to a $Z'$ and compare them with those obtained for
the canonical observables $\sigma$ and $A_{\rm FB}$ \cite{paper1}.
This analysis will be performed in a model-independent manner, 
and also includes initial state radiation (ISR) effects.
In writing down the neutral current interaction of the $Z'$ in a
model-independent way we follow \cite{paper1, Leike}.
The $Z'$ mediated amplitude for fermion
pair production in the Born approximation can be written as
\begin{eqnarray}
{\cal M}(Z')&\propto &\frac{g^2_{Z'}}{s-M^2_{Z'}}
\bigl[\bar u_e\gamma_\mu(V^e_{Z'}-A^e_{Z'}\gamma_5)u_e\bigr]
\bigl[\bar u_l\gamma^\mu(V^l_{Z'}-A^l_{Z'}\gamma_5)u_l\bigr] \nonumber \\
&=&-\frac{4\pi}{M^2_Z}
\left[\bar u_e\gamma_\mu(V_e-A_e\gamma_5)u_e\right]
\left[\bar u_l\gamma^\mu(V_l-A_l\gamma_5)u_l\right],
\end{eqnarray}
with
\begin{equation}
\label{coupl}
V_l=V^l_{Z'}\, \sqrt{\frac{g^2_{Z'}}{4\pi}\,
\frac{M^2_Z}{M^2_{Z'}-s}},
\qquad
A_l=A^l_{Z'}\, \sqrt{\frac{g^2_{Z'}}{4\pi}\,
\frac{M^2_Z}{M^2_{Z'}-s}}.
\end{equation}

It should be noted that the imaginary part of
the $Z'$ propagator is irrelevant up to LEP2 energies,
hence it is set to zero in the present analysis.

The sensitivity of observables, $\sigma_{\pm}$, 
has been assessed numerically by defining
a $\chi^2$ function as follows:
\begin{equation}
\label{Eq:chisq}
\chi^2
=\left(\frac{\Delta\sigma_{\pm}}{\delta\sigma_{\pm}}\right)^2,
\end{equation}
where $\sigma_{\pm}$ are given by Eqs.~(\ref{deltasig}) and 
(\ref{deltasigmin}), 
the uncertainty $\delta\sigma_{\pm}$ combines both statistical
and systematic errors (we take $\delta_{\rm syst}=0.5\%$ \cite{lep2}).
As a criterion to derive allowed regions for the coupling constants
if no deviations
from the SM were observed, and in this way to assess the sensitivity
of the process (\ref{Eq:leptons}) to $V_l$ and $A_l$,
we impose that $\chi^2<\chi^2_{\rm crit}$, where $\chi^2_{\rm crit}$
is a number that specifies the desired level of significance.

The observed cross section is significantly distorted in shape
and magnitude by the emission of real photons by the incoming
electron and positron.
The model predictions are corrected for ISR
effects according to \cite{Djouadi}. 
The hard photon radiation is calculated up to order $\alpha^2$ 
and the leading soft and virtual
corrections are summed to all orders by the exponentiation technique.
Each of the coefficients $F_1$ and $F_2$ is convoluted with the radiator
functions $R^e_T(k)$ and $R^e_{FB}(k)$, respectively \cite{Djouadi}, 
where $k$ is the fraction of energy lost by the radiation.
The final expression for the differential cross section is
\begin{equation}
\label{Eq:ddsigma}
\frac{\dd\sigma}{\dd\cos\theta}
=\frac{3}{8}\left[
(1+\cos^2\theta)\,\sigma_s +2\cos\theta\,\sigma_a \right],
\end{equation}
where $\theta$ is the angle between the $\mu^-$ and the $e^-$ beam
direction in the $\mu^+\mu^-$ centre-of-mass system \cite{Was}.
The symmetric and antisymmetric parts of the cross section
are given by convolutions with the ``radiators'',
\begin{equation}
\label{coeff}
\sigma_s=\int\limits_{0}^{\Delta}
\dd k\, R^e_T(k)\, \sigma_{\rm pt}(s')\, F_1(s'), \qquad
\sigma_a=\int\limits_{0}^{\Delta}
\dd k\, R^e_{FB}(k)\, \sigma_{\rm pt}(s')\, F_2(s') ,
\end{equation}
with $s'=s(1-k)$.
Due to the radiative return to the $Z$ resonance at $\sqrt{s}>M_Z$ the
energy spectrum of the radiated photons is peaked around
$E_\gamma/E_{\rm beam}\approx 1-M^2_Z/s$ \cite{Djouadi}.
In order to increase the $Z'$ signal,
events with hard photons should be eliminated from a $Z'$ search by a cut
on the photon energy, $\Delta=E_\gamma/E_{\rm beam}< 1-M^2_Z/s$.

Since the form of the corrected cross section, Eq.~(\ref{Eq:ddsigma}),
is the same as that of Eq.~(\ref{Eq:dsigma}), it follows that
the radiatively-corrected $\sigma_{\pm}$ can also be defined
by Eqs.~(\ref{sigma+}) and (\ref{sigma-}),
with the same value for $z^*$.
However, a convolution of the coefficients $F_1$ and $F_2$ with 
the radiator functions results in some shifts of the positions 
of the zeros $\sqrt{s_{\pm}}$
and the extrema in the energy dependences of $\Delta\sigma_{\pm}$.
These modifications can be kept under control by Eq.(\ref{Eq:ddsigma}).
Our numerical analysis shows that these shifts are of the order of
100~MeV.
It means that the ISR does not affect substantially 
the interference patterns shown in Figs.\ 1b and 2b.

A numerical analysis has been performed by means of the program
ZEFIT, which has to be used along with ZFITTER \cite{zfitter}.
In this way, all the SM corrections, as well as those of QED
associated with the $Z'$ contributions were taken into account.
In Fig.~3 we compare the allowed bounds on the leptonic couplings
in the ($A_l$, $V_l$) plane obtained in the improved Born 
approximation with those where we take into account also ISR effects.
The contours are obtained from two observables $\sigma_{\pm}$ and
correspond to 95\% CL ($\chi^2_{\rm crit}=6$).
According to Eq.~(\ref{Eq:Deltasig}) one can conclude that
the $\sigma_{+}$ yields ranges of observability in the
($A_l,\ V_l$) plane bounded by a circle around the origin, whereas
$\sigma_{-}$, as can be seen from
Eq. (\ref{Eq:Dsigm}), yields detectability regions that are bounded by
hyperbolas.
Fig.~3 shows the role of ISR in affecting the sensitivity to
$Z'$ parameters. It results in some relaxation of the allowed bounds 
on the parameters with respect to the improved Born predictions.
Also, a comparison of allowed bounds on ($A_l,\ V_l$) depicted in Fig.~3
with those presented in Fig.~4 of Ref.\ \cite{paper1} from
an analysis of $\sigma$ and $A_{\rm FB}$, shows
that the sensitivities of the new observables and the canonical ones 
are almost the same.

Summarizing, in this note we introduced the new observables
$\sigma_\pm$ and studied their role in getting
more direct information on $Z'$ effects compared with that
obtained from the canonical ones, $\sigma$ and $A_{\rm FB}$.
The deviations from the SM predictions, $\Delta\sigma_{\pm}$, have
very specific energy dependences which are entirely determined by the
SM parameters. In this case, one can uniquely predict the sign of
$\Delta\sigma_{\pm}$ induced by $Z'$ exchange at energies from TRISTAN to
LEP2, as well as the locations of their extrema and zeros. 
These features could be quite helpful in distinguishing the effects 
originated by a $Z'$ from those caused by other new physics sources. 
In addition, we found that at the energy points $\sqrt{s_{\pm}}$ 
($\simeq 78$ GeV and 113 GeV) both
the SM quantities $\sigma_\pm^{\rm SM}$ and their deviations
$\Delta\sigma_{\pm}$ induced by a $Z'$ tend to their minimum values 
or vanish.
These energies $\sqrt{s_{\pm}}$ are very favourable for a search for 
new physics beyond the SM and beyond $Z'$ effects.

\medskip


We would like to express our gratitude to Dr.\ A. Babich for 
helpful discussions.
This research has been supported by the Research Council of Norway.




\clearpage

\centerline{\bf Figure captions}

\vskip 15pt
\def\fig#1#2{\hangindent=.65truein \noindent \hbox to .65truein{Fig.\ #1.
\hfil}#2\vskip 2pt}

\begin{description}
\item{\bf Fig.~1}
(a) The observable $\sigma_{+}$ for muon pair production
in the improved Born approximation {\it vs.} c.m.\ energy in the SM and
in the presence of a $Z^\prime$ with mass $M_{Z^\prime}=600$~GeV
and couplings $v'{}^2+a'{}^2=1$. 
The labels 0, 1 and $-1$ correspond to the values of $v'a'$.

(b) Relative deviation of $\sigma_{+}$,
Eq.~(\ref{deltasig}), at $M_{Z^\prime}=600$~GeV.
The labels (1, 2, 3) attached to the curves correspond to
$v'{}^2+a'{}^2=1,\ 0.5,\ 0.25$, respectively. In all cases $v'a'=0$.

(c) Statistical significance $S_+$ of Eq.~(\ref{SS+}).
Parameters are as in Fig.~1a, and the integrated luminosity is
$\Lumint=300~\mbox{pb}^{-1}$.

\item{\bf Fig.~2}
(a) The observable $\sigma_{-}$ for muon pair production
in the improved Born approximation {\it vs.} c.m.\ energy in the SM and
in the presence of a $Z^\prime$ with mass $M_{Z^\prime}=600$~GeV.
Labels $V$ and $A$ correspond to $v'{}^2-a'{}^2=\pm 1$, respectively.

(b) Relative deviation of $\sigma_{-}$, 
Eq.~(\ref{deltasigmin}).
Parameters are as in Fig.~2a.

(c) Statistical significance $S_-$ of Eq.~(\ref{SS-}).
Parameters are as in Fig.~2a, and the integrated luminosity is
$\Lumint=300~\mbox{pb}^{-1}$.
Labels $V$ and $A$ correspond to $v'{}^2-a'{}^2=\pm 1$, respectively.

\item{\bf Fig.~3}
Upper bounds on the model-independent couplings ($A_l$, $V_l$) 
at 95\% CL,
in the improved Born approximation, 
as well as those also corrected for ISR.
The ``circles'' are derived from $\sigma_{+}$,
whereas the hyperbolas are derived from $\sigma_{-}$.
The energy corresponds to LEP2 with $\sqrt{s}=190$ GeV 
and $\Lumint=500~\mbox{pb}^{-1}$.

\end{description}
                  

\setcounter{figure}{0}
\begin{figure}
\begin{center}
\setlength{\unitlength}{1cm}
\begin{picture}(14.0,14.0)
\put(-1.,0.0){
\mbox{\epsfysize=14cm\epsffile{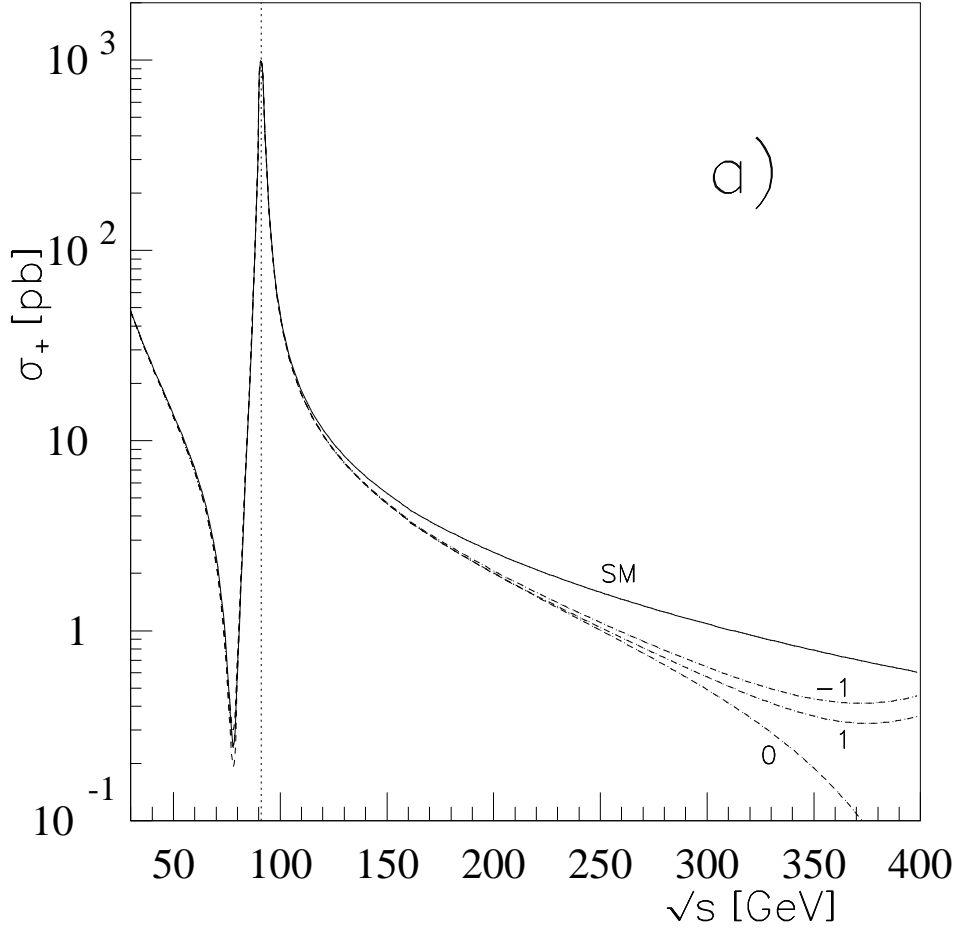}}}
\end{picture}
\caption{
(a) 
The observable $\sigma_{+}$ for muon pair production
in the improved Born approximation {\it vs.} c.m.\ energy in the SM and
in the presence of a $Z^\prime$ with mass $M_{Z^\prime}=600$~GeV
and couplings $v'{}^2+a'{}^2=1$. 
The labels 0, 1 and $-1$ correspond to the values of $v'a'$.}
\end{center}
\end{figure}
\setcounter{figure}{0}
\begin{figure}
\begin{center}
\setlength{\unitlength}{1cm}
\begin{picture}(14.0,14.0)
\put(-1.,0.0){
\mbox{\epsfysize=14cm\epsffile{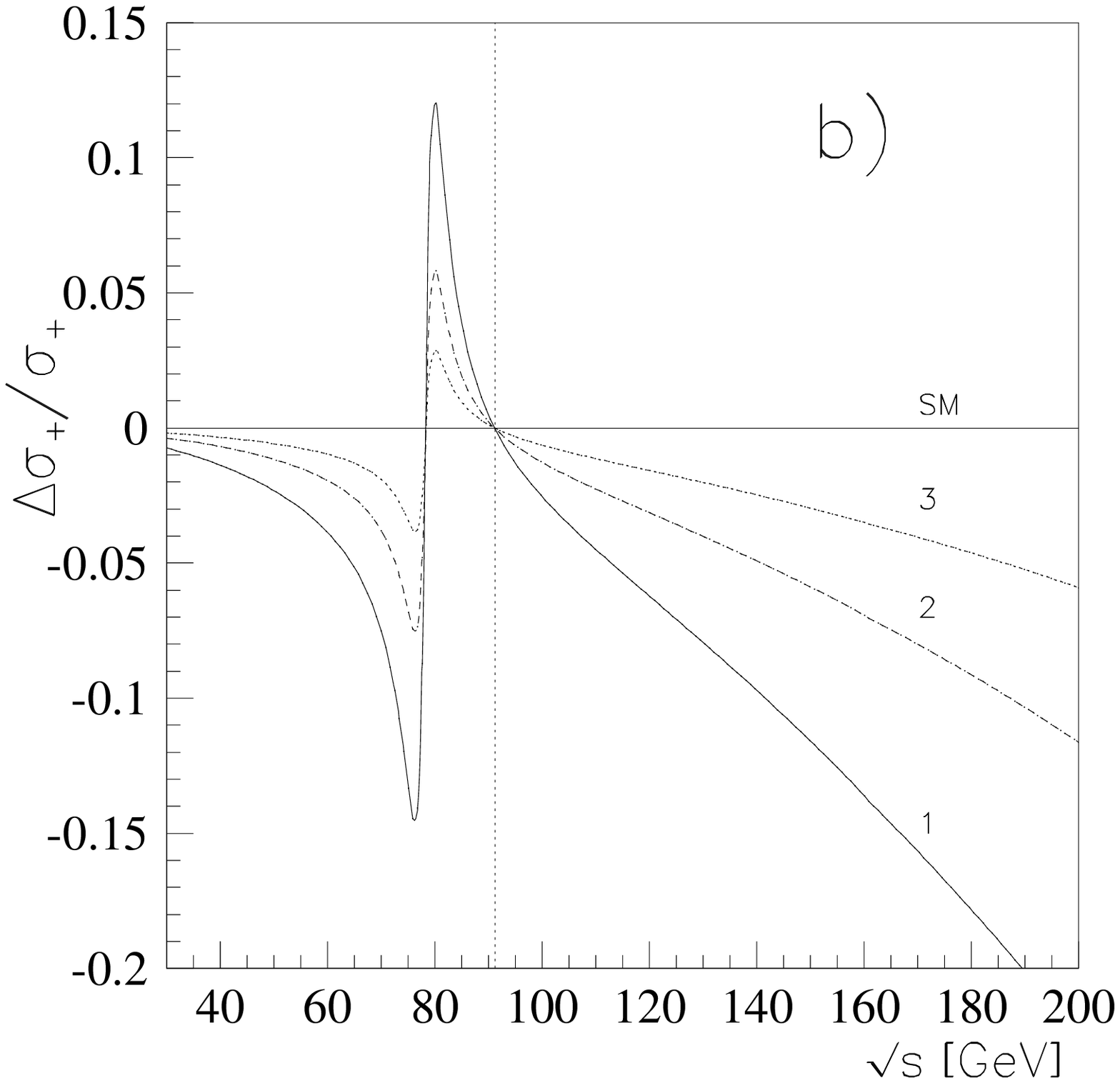}}}
\end{picture}
\caption{
(b) Relative deviation of $\sigma_{+}$,
Eq.~(\ref{deltasig}), at $M_{Z^\prime}=600$~GeV.
The labels (1, 2, 3) attached to the curves correspond to
$v'{}^2+a'{}^2=1,\ 0.5,\ 0.25$, respectively. In all cases $v'a'=0$.}
\end{center}
\end{figure}
\setcounter{figure}{0}
\begin{figure}
\begin{center}
\setlength{\unitlength}{1cm}
\begin{picture}(14.0,14.0)
\put(-1.,0.0){
\mbox{\epsfysize=12cm\epsffile{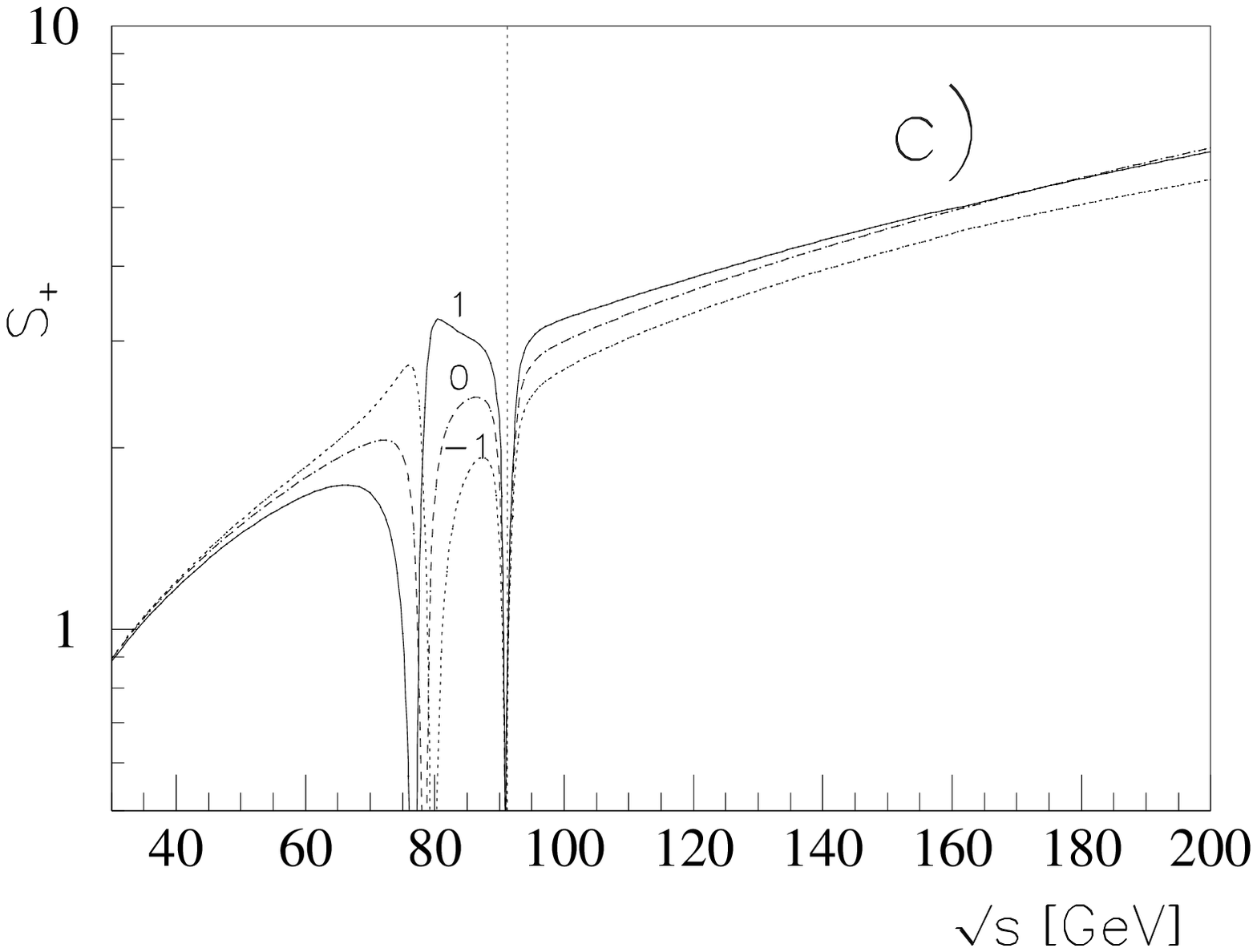}}}
\end{picture}
\caption{
(c) Statistical significance $S_+$ of Eq.~(\ref{SS+}).
Parameters are as in Fig.~1a, and the integrated luminosity is
$\Lumint=300~\mbox{pb}^{-1}$.}
\end{center}
\end{figure}
\begin{figure}
\begin{center}
\setlength{\unitlength}{1cm}
\begin{picture}(14.0,14.0)
\put(-1.,0.0){
\mbox{\epsfysize=14cm\epsffile{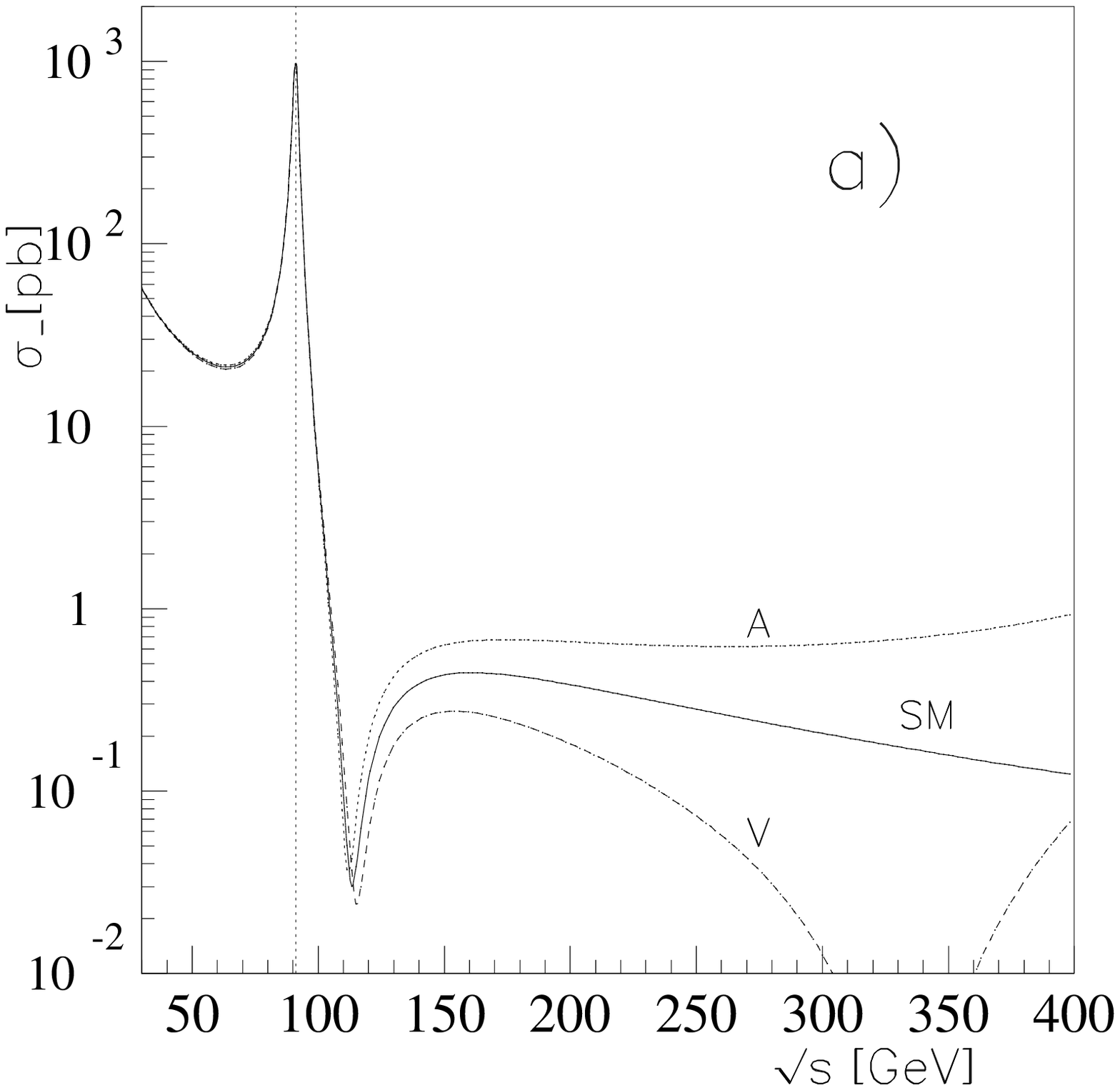}}}
\end{picture}
\caption{
(a) The observable $\sigma_{-}$ for muon pair production
in the improved Born approximation {\it vs.} c.m.\ energy in the SM and
in the presence of a $Z^\prime$ with mass $M_{Z^\prime}=600$~GeV.
Labels $V$ and $A$ correspond to $v'{}^2-a'{}^2=\pm 1$, respectively.}
\end{center}
\end{figure}
\setcounter{figure}{1}
\begin{figure}
\begin{center}
\setlength{\unitlength}{1cm}
\begin{picture}(14.0,14.0)
\put(-1.,0.0){
\mbox{\epsfysize=14cm\epsffile{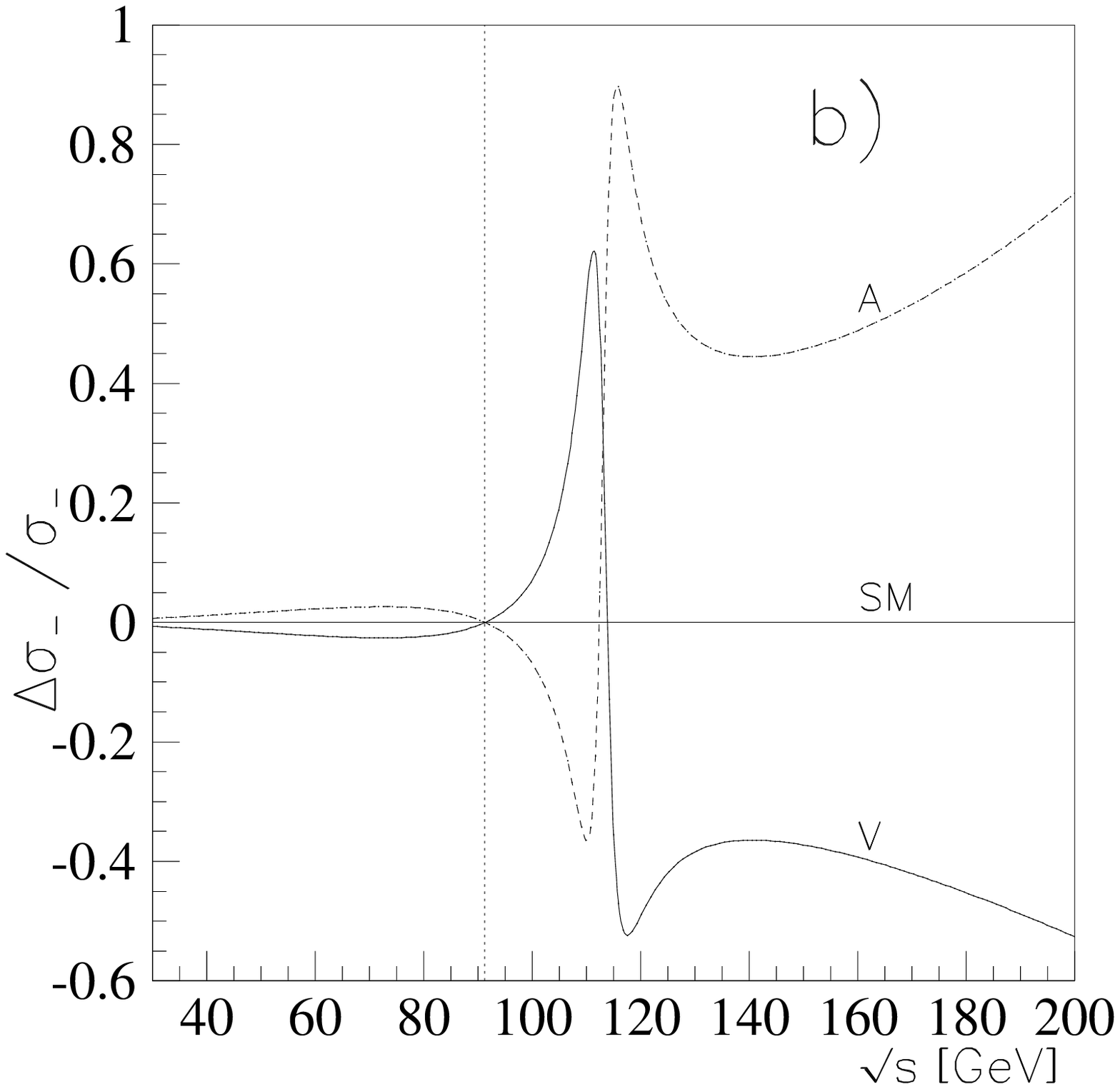}}}
\end{picture}
\caption{
(b) Relative deviation of $\sigma_{-}$, 
Eq.~(\ref{deltasigmin}).
Parameters are as in Fig.~2a.}
\end{center}
\end{figure}
\setcounter{figure}{1}
\begin{figure}
\begin{center}
\setlength{\unitlength}{1cm}
\begin{picture}(14.0,14.0)
\put(-1.,0.0){
\mbox{\epsfysize=12cm\epsffile{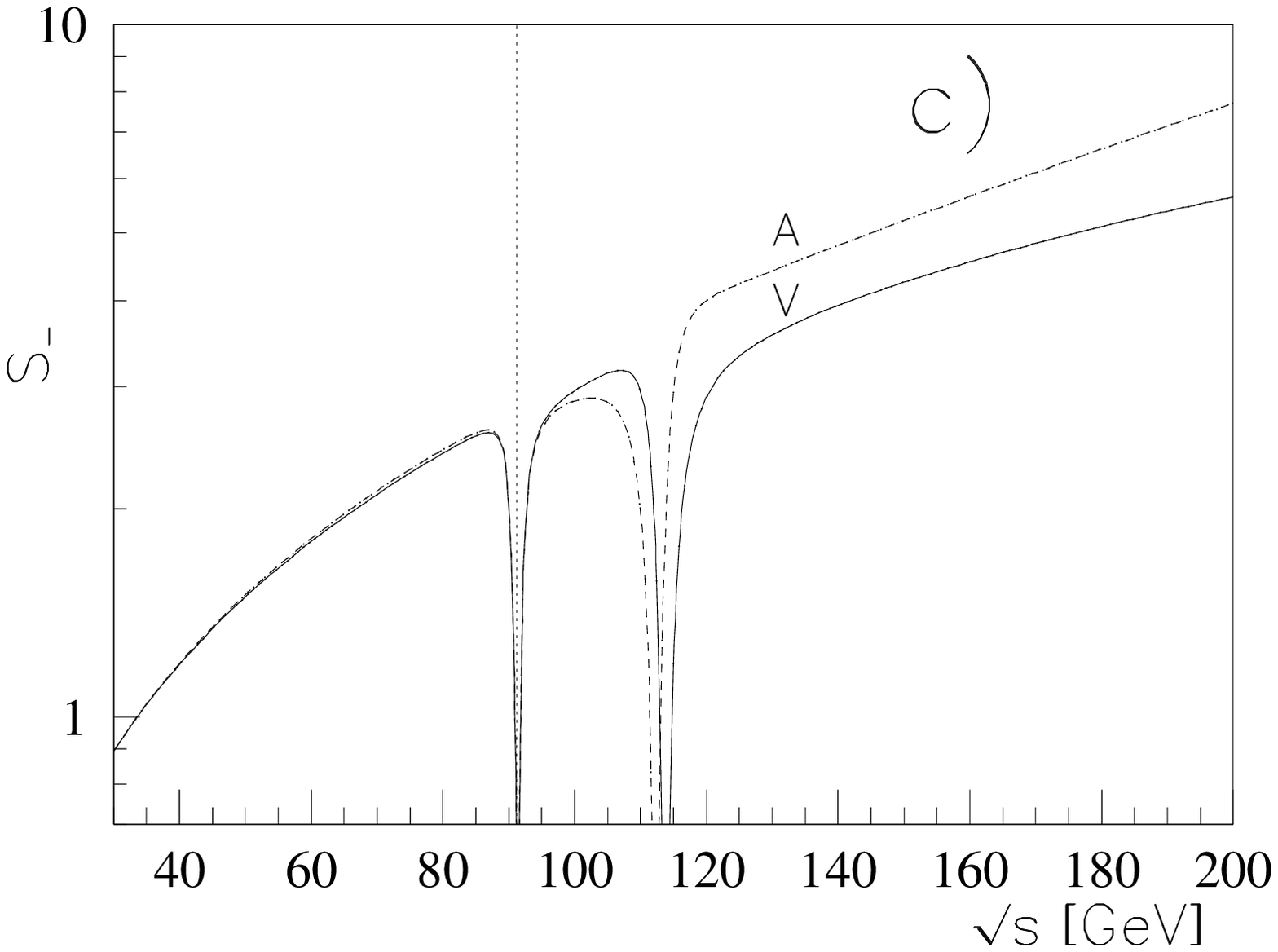}}}
\end{picture}
\caption{
(c) Statistical significance $S_-$ of Eq.~(\ref{SS-}).
Parameters are as in Fig.~2a, and the integrated luminosity is
$\Lumint=300~\mbox{pb}^{-1}$.
Labels $V$ and $A$ correspond to $v'{}^2-a'{}^2=\pm 1$, respectively.}
\end{center}
\end{figure}
\setcounter{figure}{2}
\begin{figure}
\begin{center}
\setlength{\unitlength}{1cm}
\begin{picture}(14.0,14.0)
\put(-1.,0.0){
\mbox{\epsfysize=14cm\epsffile{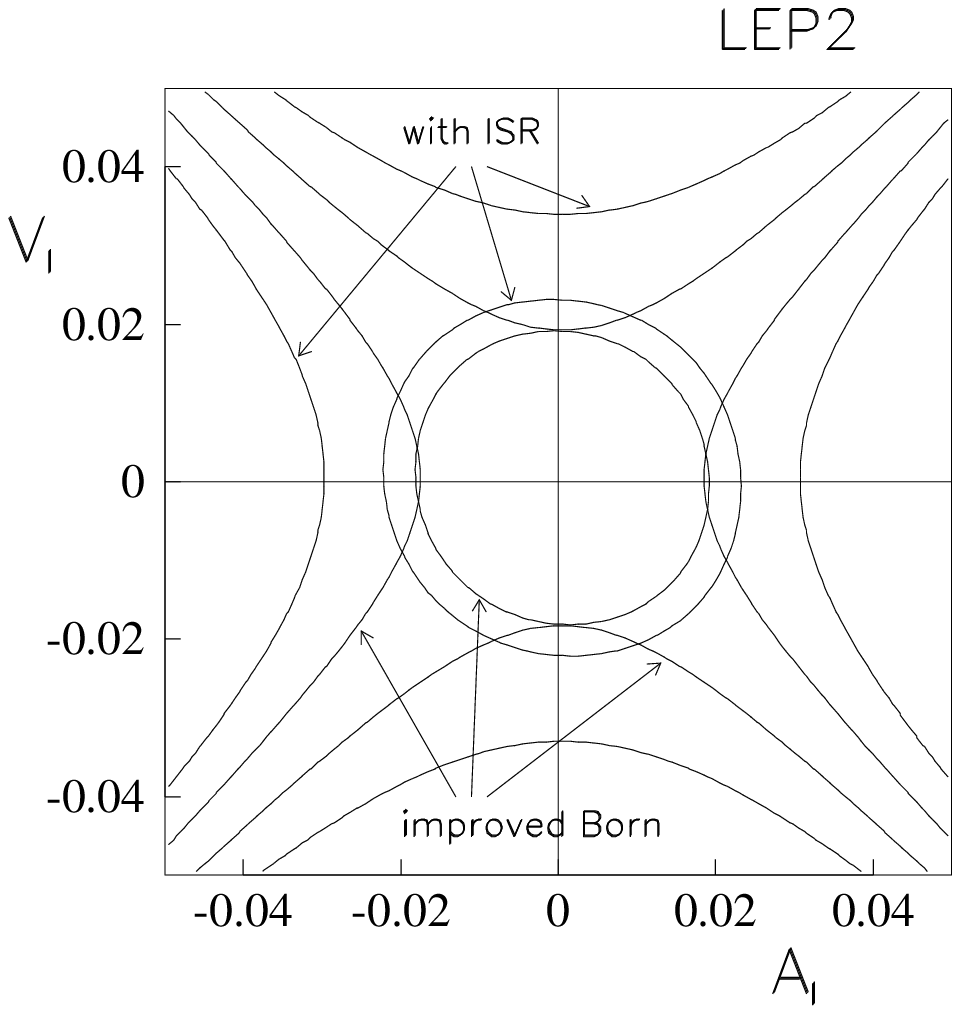}}}
\end{picture}
\caption{
Upper bounds on the model-independent couplings ($A_l$, $V_l$) 
at 95\% CL,
in the improved Born approximation, 
as well as those also corrected for ISR.
The ``circles'' are derived from $\sigma_{+}$,
whereas the hyperbolas are derived from $\sigma_{-}$.
The energy corresponds to LEP2 with $E_{\rm cm}=190$ GeV 
and ${\Lumint}=500$~{pb}${}^{-1}$.}
\end{center}
\end{figure}
\end{document}